\documentclass[conference]{IEEEtran}
\IEEEoverridecommandlockouts
\usepackage{cite}
\usepackage{amsmath,amssymb,amsfonts}
\usepackage{algorithmic}
\usepackage{graphicx}
\usepackage{textcomp}
\usepackage{xcolor}
\usepackage[acronym]{glossaries}
\newacronym{pusch}{PUSCH}{\textit{physical uplink shared channel}}
\newacronym{gnb}{gNB}{gNodeB}
\newacronym{dsp}{DSP}{digital signal processing}
\newacronym{du}{DU}{distributed unit}
\newacronym{phy}{PHY}{physical layer}
\newacronym{mac}{MAC}{multiply and accumulate operation}
\newacronym{ran}{RAN}{radio access network}
\newacronym{raw}{RAW}{read after write}
\newacronym{csr}{CSR}{control status register}
\newacronym{lut}{LUT}{look-up table}
\newacronym{fft}{FFT}{Fast Fourier Transform}
\newacronym{mmm}{MMM}{matrix-matrix multiplication}
\newacronym{nr}{NR}{new radio}
\newacronym{ipc}{IPC}{instructions per cycle}
\newacronym{mimo}{MIMO}{multiple-input multiple-output}
\newacronym{asic}{ASIC}{application-specific integrated circuit}
\newacronym{ofdm}{OFDM}{orthogonal frequency division multiplexing}
\newacronym{ue}{UE}{user equipment}
\newacronym{lsu}{LSU}{load store unit}
\newacronym{isa}{ISA}{instruction set architecture}
\newacronym{spm}{SPM}{scratchpad memory}
\newacronym{wfi}{WFI}{wait for interrupt}
\newacronym{ins}{INS}{instruction}
\newacronym{ofdma}{OFDMA}{orthogonal frequency division multiple access}
\def\BibTeX{{\rm B\kern-.05em{\sc i\kern-.025em b}\kern-.08em
    T\kern-.1667em\lower.7ex\hbox{E}\kern-.125emX}}

\begin{document}

\title{Efficient Parallelization of 5G-PUSCH on a Scalable RISC-V Many-core Processor\\
}

\author{\IEEEauthorblockN{Marco Bertuletti\\
\IEEEauthorblockA{ETH Z\"{u}rich}
Z\"{u}rich, Switzerland \\
mbertuletti@iis.ee.ethz.ch}
\and
\IEEEauthorblockN{Yichao Zhang}
\IEEEauthorblockA{ETH Z\"{u}rich}
Z\"{u}rich, Switzerland \\
yiczhang@iis.ee.ethz.ch
\and
\IEEEauthorblockN{Alessandro Vanelli-Coralli}
\IEEEauthorblockA{ETH Z\"{u}rich}
Z\"{u}rich, Switzerland \\
\IEEEauthorblockA{Universit\`a di Bologna}
Bologna, Italy \\
avanelli@iis.ee.ethz.ch
\and
\IEEEauthorblockN{Luca Benini}
\IEEEauthorblockA{ETH Z\"{u}rich}
Z\"{u}rich, Switzerland \\
\IEEEauthorblockA{Universit\`a di Bologna}
Bologna, Italy \\
lbenini@iis.ee.ethz.ch
}

\maketitle

\begin{abstract}
5G Radio access network disaggregation and softwarization pose challenges in terms of computational performance to the processing units. At the physical layer level, the baseband processing computational effort is typically offloaded to specialized hardware accelerators. However, the trend toward software-defined radio-access networks demands flexible, programmable architectures. In this paper, we explore the software design, parallelization and optimization of the key kernels of the lower \gls{phy} for \gls{pusch} reception on MemPool and TeraPool, two manycore systems having respectively 256 and 1024 small and efficient RISC-V cores with a large shared L1 data memory. \gls{pusch} processing is demanding and strictly time-constrained, it represents a challenge for the baseband processors, and it is also common to most of the uplink channels. Our analysis thus generalizes to the entire lower \gls{phy} of the uplink receiver at \gls{gnb}. Based on the evaluation of the computational effort (in multiply-accumulate operations) required by the \gls{pusch} algorithmic stages, we focus on the parallel implementation of the dominant kernels, namely fast Fourier transform, matrix-matrix multiplication, and matrix decomposition kernels for the solution of linear systems. Our optimized parallel kernels achieve respectively on MemPool and TeraPool speedups of 211, 225, 158, and 762, 880, 722, at high utilization (0.81, 0.89, 0.71, and 0.74, 0.88, 0.71), comparable a single-core serial execution, moving a step closer toward a full-software \gls{pusch} implementation.
\end{abstract}


\section{Introduction}
To provide increased flexibility, performance, and efficiency, the 5G standard foresees the introduction of novel features in its air-interface, known as \gls{nr}, such as larger bandwidths, higher spectrum frequencies, increased massive multi-user \gls{mimo}, beamforming, etc. \cite{5GNewRadio_IEEE_2019}. These enhancements require the processing of high-dimensional signals in a fraction of milliseconds. Over the last few years, a wide range of baseband processing \glspl{asic} \cite{FFTBertaccini_ASAP_2021,PengCGRAMIMO_IEEEJSC_2020,Castaneda_ESSIRC_2019} have been proposed. Industry stakeholders are, however, moving towards more flexible solutions based on \gls{ran} disaggregation and softwarization \cite{QEdge_Mavenir} to improve the time-to-market in diverse deployment scenarios.

A key direction in \gls{ran} softwarization and disaggregation is to exploit open software and hardware platforms, to ensure long-term scalability, to speed up the adoption of innovative community-developed solutions, and to reduce vendor captivity issues. The RISC-V \gls{isa} plays a strategic role in this context by enabling open software and hardware architectures and designs, without the constraints imposed by proprietary instruction sets. In this paper, we focus on the \gls{pusch} lower \gls{phy} of the uplink receiver at the \gls{gnb}  by exploring the feasibility of implementing it on MemPool \cite{Mempool_DATE_2020} and its scaled-up version TeraPool, two clusters of respectively 256 and 1024 fully programmable RISC-V cores with a shared low latency access L1 memory. The \gls{pusch} lower \gls{phy} is indeed one of the most challenging processing parts of the entire receiving chain. The main contributions of this paper are: 
\begin{itemize}{
    \item the identification of the most computationally complex kernels of \gls{pusch} lower \gls{phy};
    \item a local memory access parallel implementation of these key kernels, reducing the memory-related stalls to less than 10\% of the execution time, in MemPool and TeraPool;
    \item a flexible scheduling policy that enables executing kernels on subsets of the cluster's cores, supported by the implementation of barriers for partial group synchronization;
    \item the evaluation of the speedup of our parallel software-defined \gls{pusch} chain, compared to a single core serial execution, and of the achievable efficiency in terms of processor utilization and stall reduction. 
}
\end{itemize}
The implemented parallel kernels achieve respectively on MemPool and TeraPool speedups of 211, 225, 158, and 762, 880, 722, at utilizations 0.81, 0.89, 0.71, and 0.74, 0.88, 0.71. The speedup obtained on the whole processing chain is 871. The execution time, constrained to a realistic clock frequency of 1GHz is 0.785ms, which is close to the 0.5ms per transmission specified by the 5G PUSH standard. Our analysis thus shows that a RISC-V-based "pool of processors" architecture, whose implementation feasibility was demonstrated in \cite{Mempool_DATE_2020}, is a promising candidate for a parallel software implementation of \gls{pusch} on programmable cores. 

\section{5G PUSCH kernels complexity}
\label{complexity}
\begin{figure}[htbp]
\centerline{\includegraphics[width=\columnwidth]{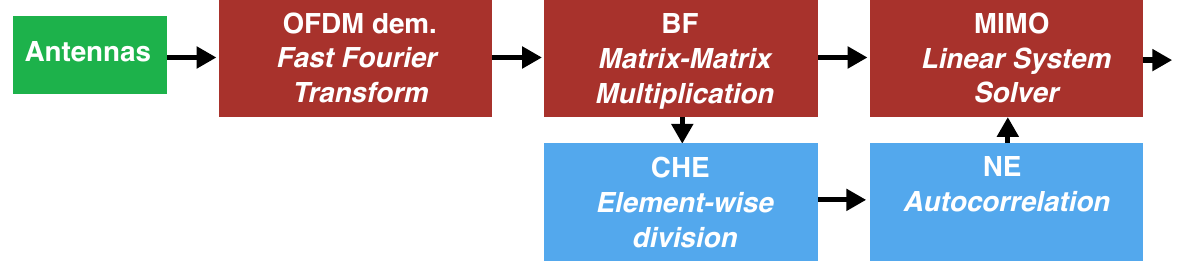}}
\caption{PUSCH processing chain steps: \acrshort{ofdm} demodulation, beamforming (BF), \gls{mimo}, channel estimation (CHE) and noise estimation (NE). The steps involving pilot symbols are reported in blue.}
\label{fig_PUSCHchain}
\end{figure}

\begin{figure}[htbp]
\centerline{\includegraphics[width=\columnwidth]{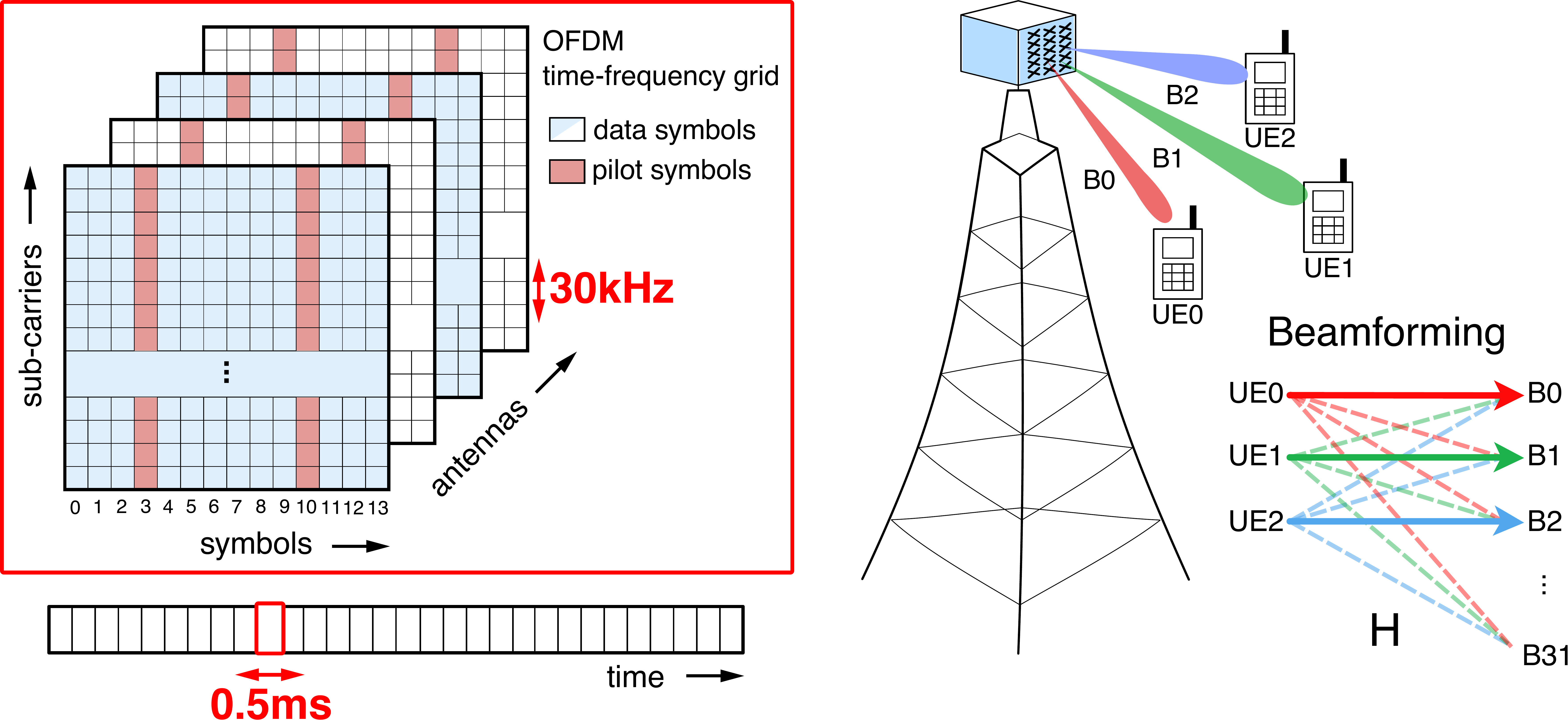}}
\caption{Time-frequency grid of an \acrshort{ofdm} system and Beamforming.}
\label{fig_OFDMgrid}
\end{figure}

This section reviews the key kernels in \gls{pusch} processing. Fig.~\ref{fig_PUSCHchain} represents the reference \gls{pusch} lower PHY receiving chain. \gls{pusch} transmission is based on \gls{ofdma} \cite{3GPP_TS}. \Glspl{ue} are multiplexed on a time and frequency grid (Fig.~\ref{fig_OFDMgrid}). Each \gls{ofdm} symbol consists of $N_{SC}$ orthogonal sub-carriers. $N_{symb}$ are sent during one slot transmission. \gls{pusch} may be interleaved in time and frequency with other channels, however, in the worst case for \gls{pusch} computational complexity the whole spectrum is allocated to this channel. \gls{ofdm} symbols are received by a set of $N_R$ antennas.

At the beginning of the baseband \gls{dsp} chain, the signal received by each antenna is translated to the frequency domain via a \gls{fft}. The complexity of this stage can be estimated as $N_{SC} \times log(N_{SC})$ complex \glspl{mac}, and the kernel is run for each antenna and each \gls{ofdm} symbol. As shown in Fig.~\ref{fig_OFDMgrid}, beamforming linearly combines the signal received by different antennas and creates $N_B$ receiving beams. This results in a \gls{mmm} with known coefficients, that requires $N_R \times N_B \times N_{SC}$ complex \glspl{mac} for each \gls{ofdm} symbol. After beamforming a $\mathbf{y} \in \mathbb{C}^{N_B}$ signal is obtained for each sub-carrier. The relation between this signal and the $\mathbf{x} \in \mathbb{C}^{N_L}$ signal transmitted by $N_L$ \glspl{ue} can be modeled as:
\begin{equation}
\mathbf{y}=\mathbf{H}\mathbf{x}+\mathbf{n}\label{eq_1}
\end{equation}
where $\mathbf{H} \in \mathbb{C}^{N_B \times N_L}$ is the channel matrix and $\mathbf{n} \in \mathbb{C}^{N_B}$ is additive white gaussian noise. In the \gls{mimo} stage, the transmitted signal is extracted from the received signal through least minimum mean squared error estimation. Before this step, the channel matrix and the noise variance are estimated. Introducing the variance of the Gaussian noise $\sigma^2$, the identity matrix $\mathbf{I}$, the estimated channel matrix $\mathbf{\hat{H}}$, its hermitian $\hat{\mathbf{H}}^H$, and the Gramian matrix $\mathbf{G}$, the \gls{mimo} stage of \gls{pusch} consists of the following:
\begin{equation}
\mathbf{x}=\left(\hat{\mathbf{H}}^H\hat{\mathbf{H}}+\sigma^2\mathbf{I}\right)^{-1}\hat{\mathbf{H}}^H\mathbf{y} = \mathbf{G}^{-1}\hat{\mathbf{H}}^H\mathbf{y}\label{eq_2}
\end{equation}
As suggested in \cite{MIMO_NorCAS_2020}, the computationally intensive matrix inversion required by \gls{mimo} can be avoided by resorting to a Cholesky decomposition of matrix $\mathbf{G}$, followed by the solution of two triangular systems. The complexity of these steps is respectively $N_L^3/3$ and $2N_L^2$, for each sub-carrier and each data \gls{ofdm} symbol. The channel matrix and the variance of noise used in \eqref{eq_2} are pilot-based estimates.
\begin{figure}[htbp]
\centerline{\includegraphics[width=\columnwidth]{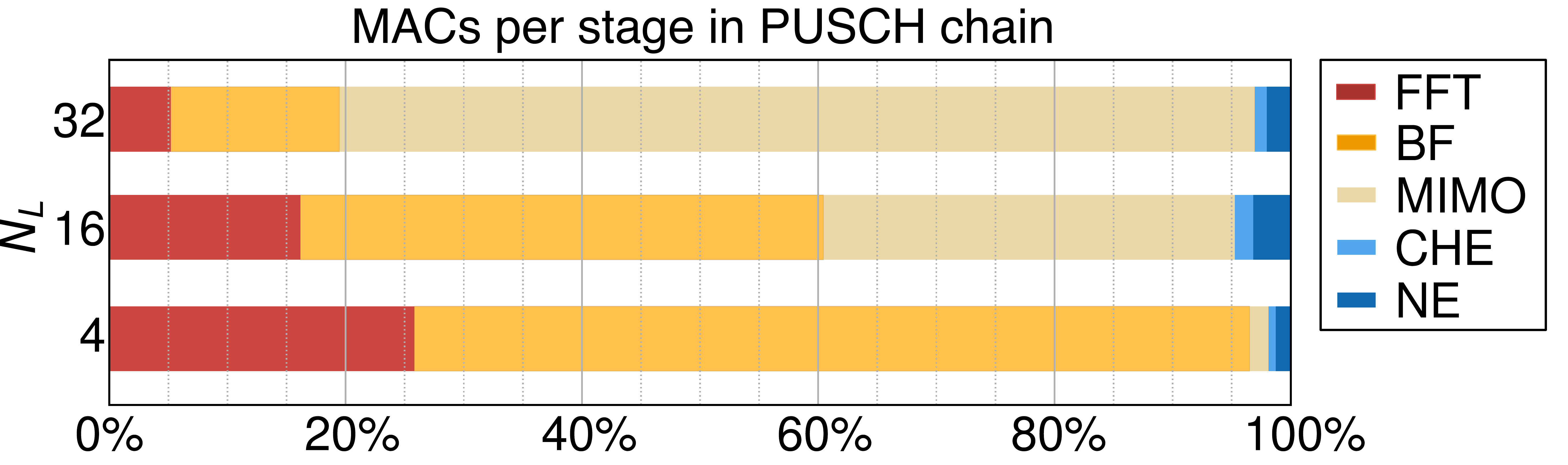}}
\caption{Complex \glspl{ofdm} allocated to each \gls{pusch} stage for different number of \glspl{ue} transmitting at the same frequency.}
\label{fig_PUSCHcomplexity}
\end{figure}
\begin{table}[htpb]
\caption{PUSCH kernels and computational complexity}
\resizebox{\columnwidth}{!}{
\begin{tabular}{|c|c|c|}
\hline
\rule{0pt}{10pt} \textbf{PUSCH stage} & \textbf{Key kernel} & \textbf{Complex MACs} \\
\hline
\rule{0pt}{10pt} OFDM dem. & Fast Fourier transform & $N_{symb} \times N_{R} \times N_{SC} \times log(N_{SC})$\\
\hline
\rule{0pt}{10pt} BF & Matrix-matrix multiplication & $N_{symb} \times N_{SC} \times N_R \times N_B$\\
\hline
\rule{0pt}{10pt} MIMO & Cholesky decomposition & $N_{data-symb} \times N_{SC} \times N_L^3/3 \times 2N_L^2$\\
\hline
\rule{0pt}{10pt} CHE & Element-wise division & $N_{pilot-symb} \times N_{SC} \times N_B \times N_L$\\
\hline
\rule{0pt}{10pt} NE & Autocorrelation & $N_{pilot-symb} \times N_{SC} \times 2N_B \times N_L$\\
\hline
\end{tabular}}
\label{tab_1}
\end{table}

In this paper, the block-type arrangement described in \cite{CHE_IEEE_2014}, is assumed and pilots are allocated to a whole \gls{ofdm} symbol, as shown in Fig.~\ref{fig_OFDMgrid}. The channel estimation block is based on least squares estimation and it consists of an element-wise matrix division.
The computational cost of this kernel is $N_B \times N_L$ \glspl{mac} for each sub-carrier and for each \gls{ofdm} symbol. The noise variance is estimated by computing the autocorrelation of the difference between the received signal and the expected transmission output, obtained from the estimated channel and the pilots. The complexity of this kernel is $2N_B \times N_L$ complex \glspl{mac} for each sub-carrier and pilot symbol.
Tab.~\ref{tab_1} reports the kernels of the \gls{pusch}, and the number of complex \glspl{mac} required for each one of them. 
\begin{figure*}[htpb]
\centering
\includegraphics[width=18cm]{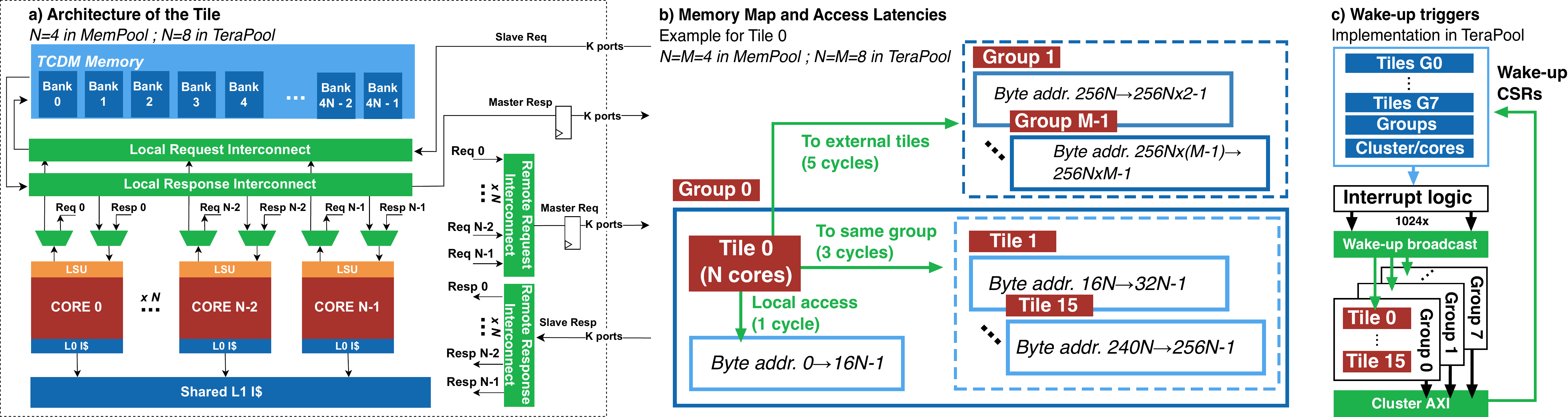}
\caption{(a) Architecture of tiles. (b) Access latency to each section of the cluster memory for the first tile. (c) Wake-up triggers in TeraPool cluster.}
\label{fig_architecture}
\end{figure*}

Let us consider a typical \gls{nr} use-case. According to the 3GPP \gls{nr} numerology, we consider a bandwidth of 100MHz, with sub-carrier spacing 30KHz, corresponding to 3276 sub-carriers. We assume 14 symbols per transmission and 2 pilot symbols, 64 receiving antennas, and 32 beams. Fig.~\ref{fig_PUSCHcomplexity} represents the complexity allocated to each kernel of the \gls{pusch} processing chain as a percentage fraction of the total. Most of the effort is in \gls{ofdm} demodulation and beamforming stages, the impact of \gls{mimo} stage depends on the number of \glspl{ue} involved. According to Amdahl's law, this analysis shows that the throughput of the chain would greatly benefit from the speedup of \gls{fft}, \gls{mmm}, and Cholesky decomposition.
\section{MemPool-TeraPool architecture}
In this section, we present the hierarchical architecture of TeraPool, a general-purpose compute cluster extended from the scalable many-core architecture of MemPool\cite{Mempool_DATE_2020}. The clusters compute unit is \textit{Snitch} \cite{Snitch_IEEE_2021}, a single-stage 32-bit RISC-V core supporting the RV32IMAFD custom extensible \gls{isa} \footnote{In this paper we do not discuss PUSH specific ISA extensions: this step is left for future work.}. Instructions whose execution requires more than one cycle are offloaded to pipelined functional units through a dedicated port. The \gls{lsu} handles memory transactions and issues up to 8 outstanding loads and stores, hiding the L1 interconnect latency. Fig.~\ref{fig_architecture} (a) shows the architecture of a \textit{tile}\cite{Mempool_DATE_2020}, which is the first building block allowing massive replication. In MemPool each tile contains 4 Snitch cores, sharing 2 KiB of L1 instruction cache and 16 banks, with 1 KiB each of local L1 data memory. Similarly, in TeraPool a tile has 8 Snitch cores with 4 KiB instruction cache and 32 banks, with 1 KiB each of local L1 data memory. Each core in a tile accesses the local memory in one cycle, through a fully connected interconnection.

The next hierarchy level is the \textit{group}\cite{Mempool_DATE_2020}. In both MemPool and TeraPool, each group has 16 tiles. The main bottleneck in a large shared memory many-core architecture is interconnection routing. A tile-to-tile crossbar though the whole cluster would not allow physical feasibility. A 16 x 16 fully connected crossbar is thus restricted to the group level. Each tile is connected with the K groups of other tiles in the cluster through master request and slave response ports, that are used for access to remote memory banks in the same local group within 3 cycles and to remote groups in 5 cycles, as Fig.~\ref{fig_architecture} (b) shows. Overall, MemPool has 256 Snitch cores, 4 groups, and 1024 1 KiB banks of L1 memory, equal to 1 MiB of SRAM. Similarly, TeraPool has 1024 cores, 8 groups, and 4096 1 KiB banks of L1 memory, equal to 4 MiB of SRAM. We do not discuss here physical implementation strategies for Terapool and Mempool, as our focus is on software design and optimization. The interested reader is referred to \cite{Mempool_DATE_2020}.

\section{Programming model and synchronization}
In this section, we present the fork-join programming model adopted for the parallel execution of the \gls{pusch} kernels. The sequential execution of the \gls{pusch} kernels is split into portions without data dependencies, that are executed in parallel over multiple cores of the cluster. At the end of a parallel task, cores are synchronized, ensuring the consistent write-back of the results.
To execute the kernels on a subset of cores, we implement partial synchronization barriers. 

When a kernel runs on the whole cluster, the cores ending a parallel task atomically increment a barrier variable and enter a \gls{wfi} sleep state. The last core incrementing the barrier variable writes in a wake-up \gls{csr} of the system, and activates a broadcasted wake-up trigger, waking up all the cores, as shown in Fig.~\ref{fig_architecture} (c). A core can also selectively wake up another one, writing its ID in the wake-up \gls{csr}. This allows to synchronize a subset of cores, but the last core completing the parallel task must individually wake up the processing elements involved in the computation. To simultaneously assert a subset of the wake-up triggers, we add one \gls{csr} to selectively wake up groups and one \gls{csr} per group to selectively wake up its tiles. Enabling the wake-up of a subset of cores allows to introduce fast partial synchronization barriers. When a kernel runs in parallel on a subset of cores, the cores terminating the execution of the parallel task increment a barrier variable in their local memory. The last core completing the task sends wake-up triggers with different granularity, depending on the total number of cores involved.

\section{Implemented kernels}
According to Amdahl's law, the key kernels that must be efficiently parallelized to boost the throughput of the \gls{pusch} processing chain are \gls{fft}, \gls{mmm}, and matrix decomposition. These kernels are implemented assuming that the input and output data reside in L1 memory and their parallelization targets the multi-banked memory structure of MemPool and TeraPool clusters. In such a large interconnected memory, contentions may occur when two cores in the same tile access the same local bank, or the same remote group. This generates stalls of the \gls{lsu} and increases the access latency. The problem is addressed, and contentions are avoided by carefully placing the data structures in memory, emphasizing local loads and stores. When local data access is impossible, the access pattern of the cores can be rearranged to avoid simultaneous access to the same group from cores in the same tile. In the following subsections, the parallel implementation of the most computationally intensive \gls{pusch} kernels is described.

\subsection{Fast Fourier transform}
\begin{figure}[htbp]
\centerline{\includegraphics[width=\columnwidth]{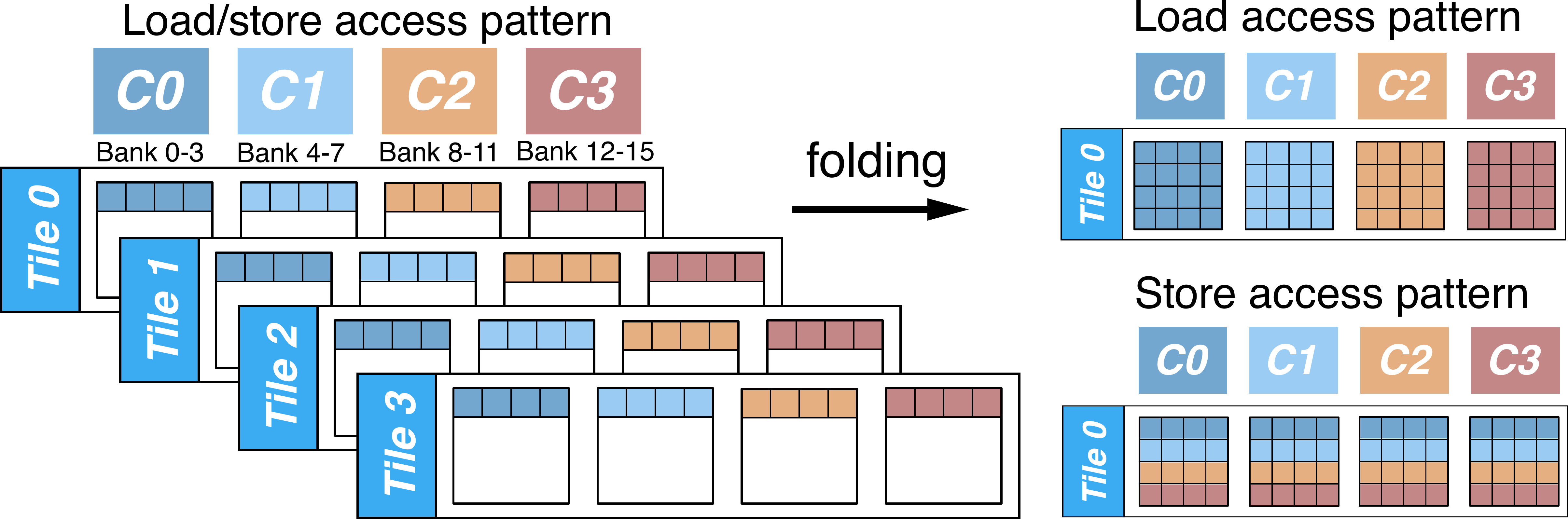}}
\caption{\gls{fft} folded access pattern for a 64-points FFT.}
\label{fig_FFTaccess}
\end{figure}

We chose a radix-4 decimation in frequency Cooley-Turkey \gls{fft} approach. The radix is chosen to ease the memory accesses in local banks of MemPool and TeraPool, where each core has 4 local banks. In the $k^{th}$ stage of an N-points \gls{fft}, the radix-4 butterfly gets 4 inputs at a distance $N/(4 \times 4k)$. Each core computes 4 butterflies. For a 64-points \gls{fft} the accessed elements are reported in different colours in Fig.~\ref{fig_FFTaccess}. Since the input vector unrolls over the whole memory, the access to 3 out of 4 elements will likely be external and generate conflicts. The input vector is thus folded in the local banks so that each set of the $N/(4 \times 4k)$ inputs is stored in a memory row. At the end of the computation, each core stores the results with the same folding scheme in the local banks of cores that are using them in the following \gls{fft} stage.
The stage-by-stage division of the \gls{fft} in smaller \glspl{fft} computed over a sub-set of cores helps reduce the synchronization overhead because only the cores producing the inputs of the same \gls{fft} for the following stage need to be synchronized. Depending on the size of the input vector only $N/4$ cores are used. The rest of the cores in the cluster are allocated to the computation of other \glspl{fft} of the same size: MemPool fits $(256 \times 4) / N$ \glspl{fft} and TeraPool $(1024 \times 4) / N$ \glspl{fft}. Cores working on different \glspl{fft} are independently synchronized.
\begin{figure}[htbp]
\centerline{\includegraphics[width=\columnwidth]{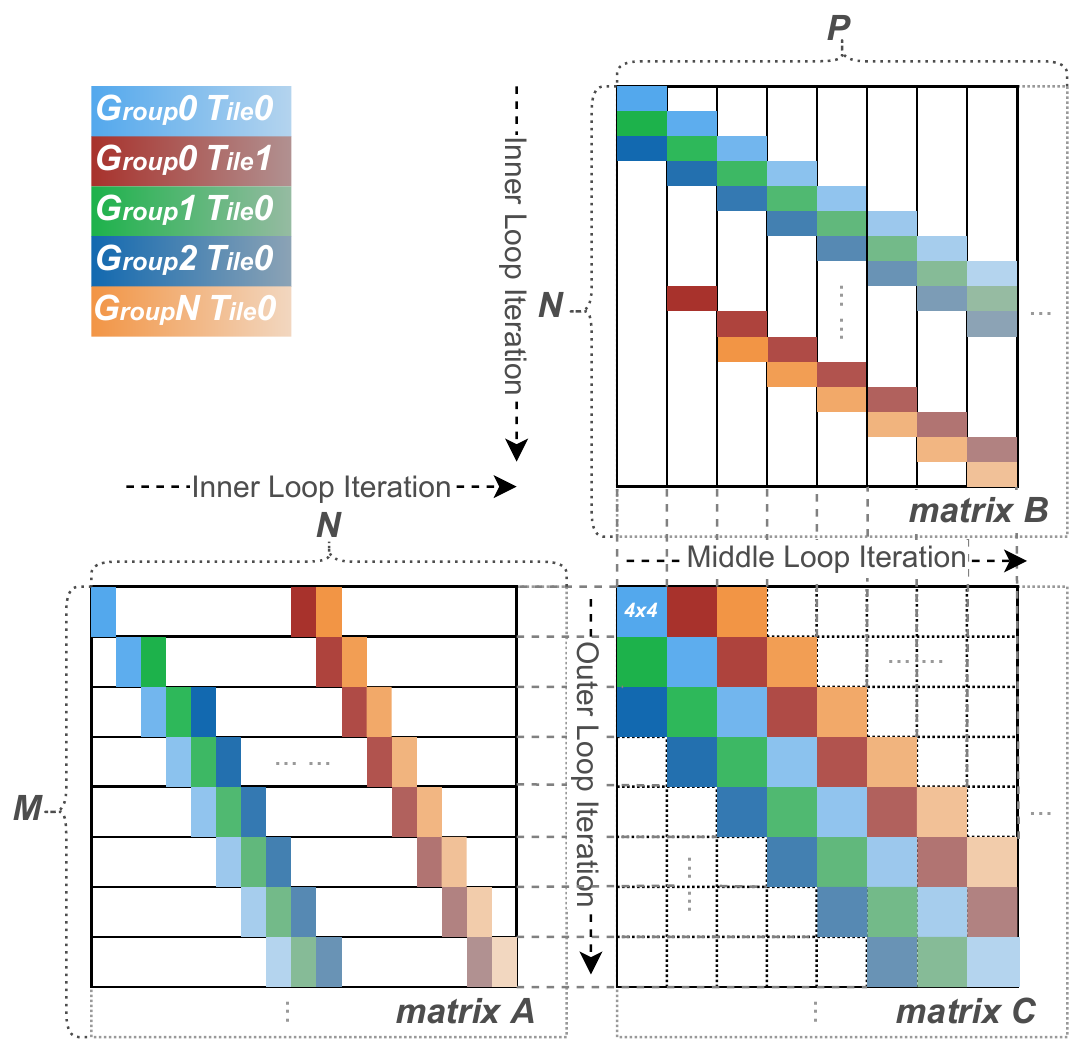}
}
\caption{Kernel of \gls{mmm} with 4x4 compute-window based optimization.}
\label{fig_AlgorithmMatMul}
\end{figure}
\begin{figure}[htbp]
\centerline{\includegraphics[width=\columnwidth]{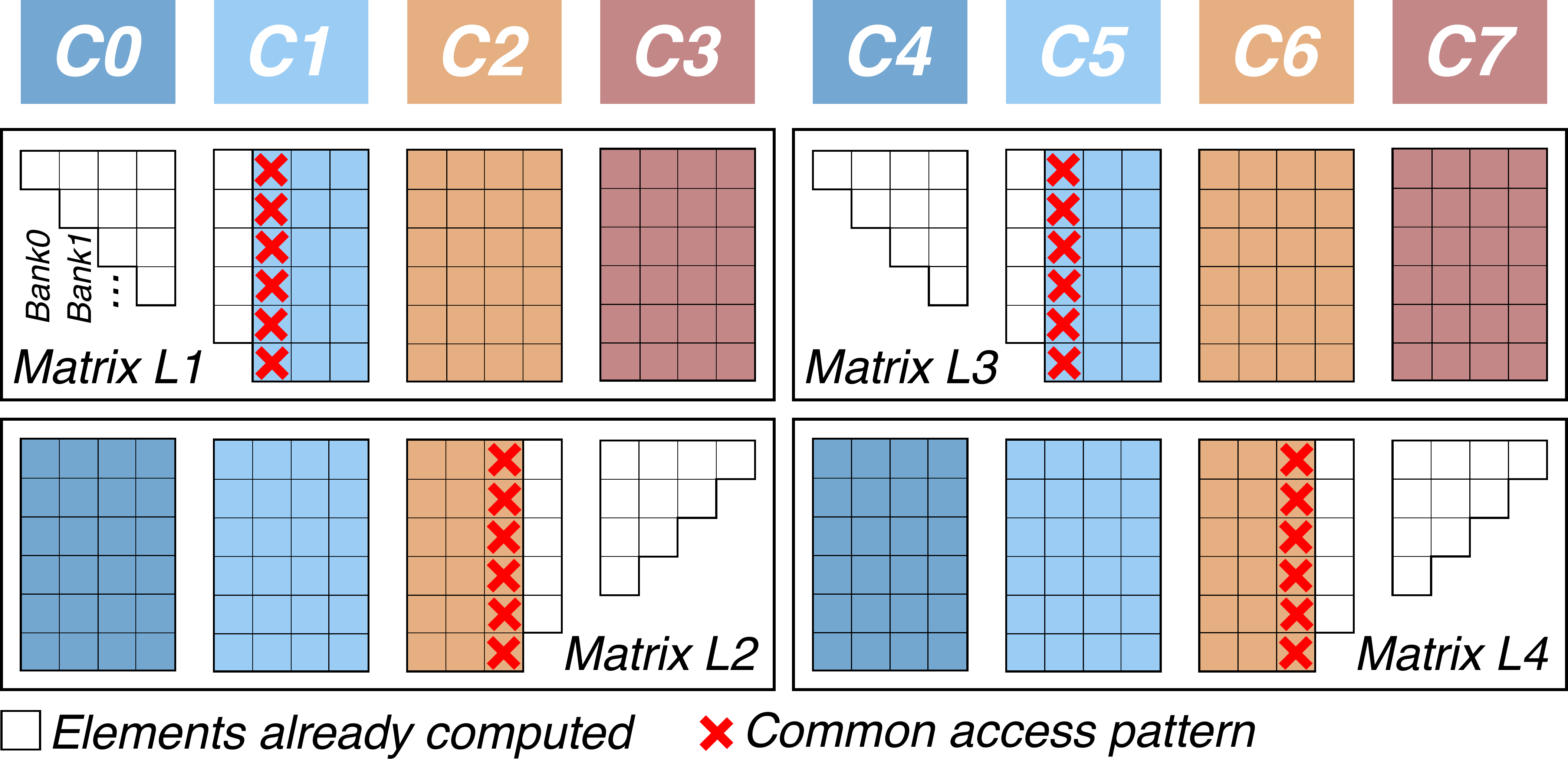}}
\caption{Cholesky decomposition parallelization scheme and replication over the cluster for a 16x16 matrix.}
\label{fig_Choleskyaccess}
\end{figure}
\subsection{Matrix-matrix multiplication}
To efficiently handle the \gls{mmm} on our architecture, 4x4 windows of the output matrix are computed at a time for two reasons. First, we achieve maximum utilization of the register files in Snitch, using all the 30 registers available for programming purposes in its \gls{isa}: 8 registers for inputs, 16 for the accumulation of temporary results, 3 for address increment, and 3 for loop control. Second, the large window size increase data reuse. Computing a 4x4 window requires 8 loads of 32-bit words per 16 \glspl{mac}. This memory accesses vs. computing operations ratio is lower than respectively 12 or 16 memory loads per 16 \glspl{mac}, required in a 4x2 or a 2x2 window kernel. 
The parallelization scheme of the implemented kernel is represented in Fig.~\ref{fig_AlgorithmMatMul}. An \textit{M}x\textit{N} matrix \textbf{A} and an \textit{N}x\textit{P} matrix \textbf{B} are multiplied to obtain an \textit{M}x\textit{P} matrix \textbf{C}. The kernel consists of three loops, where cores span over the whole input matrices rows and columns to compute an entire output window without reductions. In the outer loop, each core is allocated 4 rows of matrix \textbf{A}. 

\begin{figure*}[htpb]
\centering
\includegraphics[width=18cm]{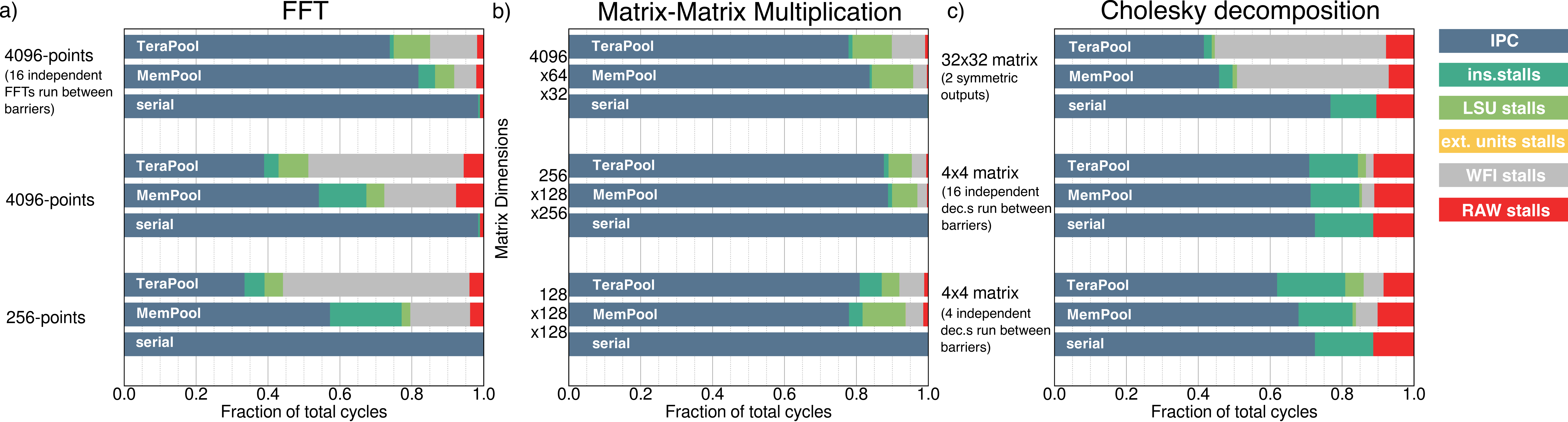}
\caption{Fraction of instructions and stalls over the total cycles for the \gls{pusch} kernels.}
\label{fig_IPC}
\end{figure*}

To maximize the utilization of cores in the cluster, cores from different tiles can operate on the same group of four rows, to generate different windows in the output matrix. A conflict occurs when cores in the same tile access data in the same group. To avoid this and fully utilize the ports for external accesses, cores from the same tile are forced to work on rows, whose elements are located in different groups.
In the middle loop, a core assigned to a row of matrix \textbf{A} spans over multiple groups of 4 columns of matrix \textbf{B}, to complete the computation of multiple output 4x4 windows. If cores of the same tile generate a bank conflict, the loop starting point for one of the cores shifts to the neighbouring four columns, and round-robins back to complete the loop.
The inner loop specializes in the computation of the 4x4 output window. To avoid conflicts in accesses to the same locations of the output matrix, the cores working on the same output window shift their starting point on both rows and columns, then they round-robin back.
\subsection{Cholesky decomposition}
The Cholesky decomposition of matrix $\textbf{G}$ in the lower triangular and upper triangular matrices $\textbf{L}$ and $\textbf{L}^H$ follows Cholesky-Crout algorithm, which computes the output matrix $\textbf{L}$ column by column. At each iteration, a new column of matrix $\textbf{L}$ is generated, and all the already computed elements of a row must be accessed to produce the new row element on this column. In the parallel implementation, each core computes 4 rows in the output matrix. To avoid conflicts in the access to the elements of a row, the output matrix is folded in memory, rows are stored in the same bank. The staircase pattern of the kernel allocates more computations to the cores accessing the bottom rows of the output lower triangular matrix, unbalancing the workload and increasing the synchronization overhead. We thus replicate two instances of the kernel with different input matrices and mirrored outputs, as represented in Fig.~\ref{fig_Choleskyaccess}. Depending on the input matrix size, a different number of cores is used in the fine-grained parallelization. The remaining cores in the cluster can work on the decomposition of other matrices and be independently synchronized.
\section{Results}
Fig.~\ref{fig_IPC} represents the \gls{ipc} of a \textit{serial} implementation of the kernels run on a single TeraPool core and the average \gls{ipc} for the parallel implementations running on MemPool and TeraPool. We also represent a breakdown of idle time due to synchronization (\gls{wfi} stalls) or architectural stalls: instruction stalls, \gls{lsu} stalls, stalls of the external pipelined units, and \gls{raw} stalls. The latter originate when the register file of Snitch must wait for the output of the multiplication and division unit and the \gls{lsu}.
In Fig.~\ref{fig_IPC} (a), the parallel implementations take into account the replication of independent \glspl{fft}, to employ all the cores of a cluster: MemPool fits 16 256-points \glspl{fft} and 1 4096-points \gls{fft}, TeraPool fits 64 256-points \glspl{fft} and 4 4096-points \glspl{fft}. For larger input vectors the impact of synchronization overhead is reduced because there are fewer groups of cores simultaneously writing in the system \glspl{csr} to trigger an interrupt. Having a larger cluster, TeraPool has a larger fraction of \gls{wfi} stalls with respect to MemPool. The same stage of different independent \glspl{fft} is run between the synchronization barriers to reduce the synchronization overhead. Running 16 independent 4096-points \glspl{fft} between the barriers we obtain 16 \glspl{fft} on MemPool and 64 \glspl{fft} on TeraPool. The \gls{ipc} is increased to respectively 0.82 and 0.74.

The \gls{ipc} of the single core and parallel implementations of \gls{mmm} for different input dimensions are shown in Fig.~\ref{fig_IPC} (b). As for the \gls{fft}, the kernel implemented on TeraPool exhibits more \gls{wfi} stalls. Since the same kernel is parallelized over the entire cluster, cores in the larger TeraPool configuration get fewer instructions. This increases the fraction of instruction stalls. The few leftover \gls{lsu} stalls are caused by conflicts from the cross-accesses in the two input matrices. The relative fraction of these stalls is smaller in TeraPool than in MemPool, because the \gls{lsu} stalls are hidden by the overlapping instruction stalls. For the 256x128x256 problem, MemPool achieves 0.89 \gls{ipc} and TeraPool 0.88 \gls{ipc}, which leads respectively to 145 and 558 \glspl{mac}/cycle.

The example use-case described in section \ref{complexity} requires a 4096x64x32 \gls{mmm}. In this case the irregular matrix shape unbalances the workload assigned to the cores, nevertheless MemPool achieves 0.84 \gls{ipc} and 134 \glspl{mac}/cycle TeraPool achieves 0.78 \gls{ipc} and 487 \glspl{mac}/cycle.

As shown in Fig.~\ref{fig_IPC} (c), both the single-core and the parallel versions of the Cholesky decomposition kernel are influenced by its \textit{staircase} structure. The inner loops of the algorithm count a different number of elements at each column iteration, making it difficult to hide the \gls{raw} stalls on data produced by the multiplication and division units. The cores working on the matrix central rows are assigned a smaller workload and conclude their task in advance, increasing the synchronization overhead. In the parallel implementations, independent decompositions are replicated to fit all the banks of a cluster. Respectively on MemPool and TeraPool we can fit 256 and 1024 single-core decompositions of 4x4 matrices, 32 and 128 fine-grained parallel decompositions on couples of 32x32 symmetric output matrices. 

The same subset of cores can also generate multiple decompositions on independent inputs before the barrier to reduce the synchronization overhead. Using this strategy and running respectively 16x256 and 16x1024 single core Cholesky decompositions of 4x4 matrices on the whole MemPool and TeraPool clusters, we achieve an \gls{ipc} of 0.71.

\begin{figure}[htpb]
\centerline{\includegraphics[width=\columnwidth]{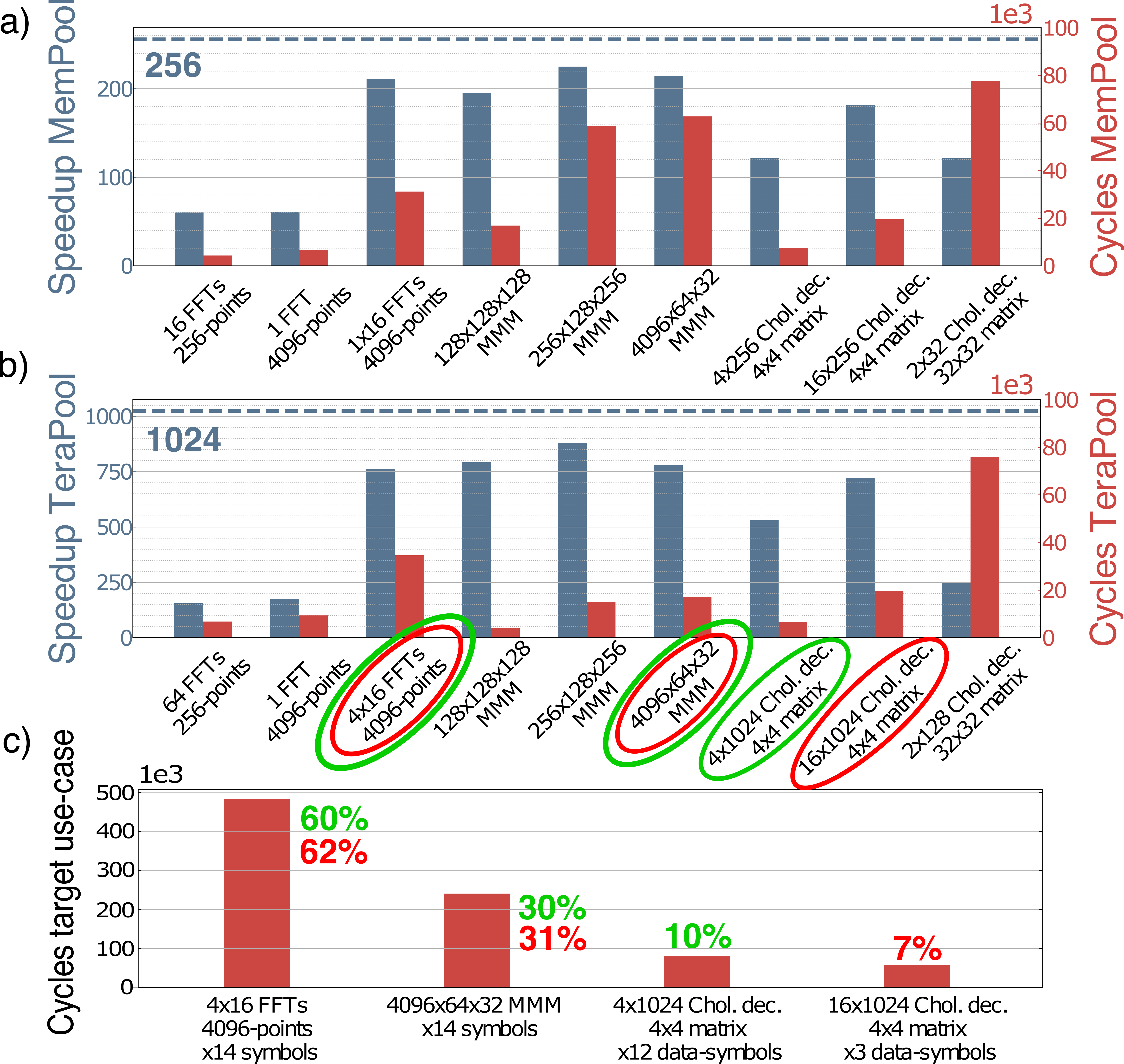}}
\caption{(a-b) Speedup with respect to a serial single core execution, and total number of execution cycles. The use-case benchmarks are circled in green and red. (c) Total number of cycles per use-case kernel and percentage over the total execution time.}
\label{fig_speedup}
\end{figure}

Fig.~\ref{fig_speedup} (a) and (b) represent the speedup of the parallel implementations, with respect to a serial single core execution, and the total execution time of the kernels. The theoretical limit, corresponding to the total number of cores used, is represented as a blue dotted line. Considering 4 \glspl{ue} active on the same frequency, the use-case described in section \ref{complexity} is addressed executing 64 4096-points \glspl{fft} and a 4096x64x32 \gls{mmm} for each of the 14 \gls{ofdm} symbols, 4096 Cholesky decompositions of 4x4 matrices for 12 data symbols. The overall speedup that can be achieved on TeraPool, using the kernels circled in green in Fig.~\ref{fig_speedup} (b) is 848, the corresponding execution cycles and their percentage over the total is represented in Fig.~\ref{fig_speedup} (c). If 4x4096 Cholesky decompositions of 4x4 matrices are scheduled every 4 data symbols, the \gls{ipc} of the last stage increases. The overall speedup obtained using the kernels circled in red in Fig.~\ref{fig_speedup} (b) is 871. 
The total execution cycles is shown in Fig.~\ref{fig_speedup} (c) with a breakdown on the kernels. Executing the full \gls{pusch} then requires 785 thousand cycles, corresponding to 0.785ms when the cluster runs ad 1GHz. An implementation analysis of the MemPool Architecture \cite{Mempool_DATE_2020} demonstrates that this speed is achievable in FINFET technology (12nm and beyond), assuming a speedup of 30\% with respect of the less advanced 22nm FDSOI technology used in \cite{Mempool_DATE_2020}. The 0.5ms timing constraint for one transmission in 5G PUSH can be met with customization of the RISC-V cores with domain-specific instructions (e.g. FFT butterfly), which will be explored in future work.

\section{Conclusions}
In this paper, we proved the flexibility of MemPool and TeraPool many-core architectures, leveraging their shared memory structure in the parallelization of the key kernels of the lower \gls{phy} receiving chain of 5G \gls{pusch}. The efficient parallelization of kernels with different memory access patterns on architectures with a large shared data memory was achieved. It was demonstrated that parallelizing the most computationally complex kernels in \gls{pusch} processing, namely \gls{fft}, \gls{mmm}, and Cholesky decomposition, TeraPool can provide speedup up to 871 with respect to the serial execution on a single RISCV core. The total execution time, at a realistic clock frequency of 1GHz, is 0.785ms, which can be further improved toward the 0.5ms target execution time by implementing domain-specific instruction extension in the RISC-V cores. Our work represents a first concrete step toward a 5G software-defined \gls{ran} over a fully open-source parallel RISC-V architecture. 

\bibliographystyle{IEEEtran} 
\bibliography{bibliography.bib}


\end{document}